\documentclass[aps,twocolumn,showkeys,amssymb]{revtex4-1}
\date{\today}
\usepackage{graphicx}
\usepackage{color}

\usepackage{array}
\usepackage{graphicx}
\usepackage{amssymb}
\usepackage{longtable}

\usepackage{bm}
\begin{document}

\title{Shell evolution and its indication on the isospin dependence of the spin-orbit splitting}
\author{Zhen-Xiang Xu}
\author{Chong Qi}
\thanks{Email: chongq@kth.se}
\affiliation{KTH (Royal Institute of Technology), Alba Nova University Center,
SE-10691 Stockholm, Sweden}\date{\today}

\begin{abstract}
The available experimental data on shell evolution indicate that the strength of the spin-orbit single-particle potential may be enhanced in neutron-rich nuclei. We observe that such a simple scheme destroys the harmonic oscillator magic numbers $N=8$ and 20 and generates new magic numbers like $N=6$, $14$, 16, 32 and 34. The traditional magic numbers like $N=28$ and 50 and $N=14$ seen in $^{22}$O are eroded in neutron-rich nuclei due to the sensitivity of larger-$l$ orbitals to the depth of the central potential but they are more robust than the harmonic oscillator magic numbers. The $N=82$ shell closure persists in neutron-rich nuclei while the previously proposed shell closures like $N=40$ and 70 do not emerge.
Both mechanisms contribute to enhancing the $N=56$ and 90 gaps by splitting the $1d_{5/2}$ and $0g_{7/2}$ and the $0h_{9/2}$ and $1f_{7/2}$ orbitals. 
\end{abstract}

\keywords{Shell evolution; Woods-Saxon potential; Spin-orbit coupling; Isospin dependence}

\maketitle


One of the most important and challenging frontiers of nuclear structure physics is the study of nuclei at the limit of stability, especially neutron-rich nuclei with weakly bound neutrons. A topic of particular interest is the evolution of the shell structure in those nuclei. That is, the magic number may change dramatically depending on the $N/Z$ ratio when we move towards the particle drip lines \cite{Sor08}. Such study is important not only due to the expected variation in properties of nuclei and the formation of island of inversion but also for the understanding of nuclear astrophysics as well as the nucleon-nucleon interaction. Nowadays it is rather commonly accepted that the $N=8$ and 20 Harmonic Oscillator (HO) shell closures disappear in neutron-rich nuclei \cite{Sor08}. On the other hand, new magic numbers like $N=14$, 16, and $32$ may emerge. Shell model calculations suggest that $N=34$ may also be a magic number in Ca isotopes, depending on the effective interactions used (see, e.g., Fig. 2 in Ref. \cite{Holt12}). 

Among the features intensively discussed on the mechanisms behind the  shell evolution phenomena, one can mention the effects induced by the tensor components of the two-body shell model interaction as well as the three-body interactions (see Refs. \cite{Sor08,Holt12} and references therein). 
In the shell model scheme, it is usually assumed that the evolution of the shells are solely determined by the correlation between valence nucleons in the open shell. The shell evolution has also been analyzed from a self-consistent mean-field point of view without assuming an inert core (see, e.g., Ref. \cite{Sor08,Zal09}). This study suggests that a significant part of the spin-orbit splitting may come from the two-body spin-orbit (SO) and tensor forces and three-body forces. The availability of experimental data in nuclei with large $N/Z$ ratios may provide a ground to constrain the properties of different components of the interaction, such as the isovector channel of the SO interaction, which are not well defined but may be responsible for the shell evolution.

The two-nucleon separation energies and excitation energies of $2^+_1$ states in even-even nuclei, where a jump may have its origin in shell closure, are often used as possible signatures for the evolution of shell structure within a given isotopic or isotonic chain.
In the idealized Hartree-Fock case, the one- and two-nucleon separation energies are simply related to the energy of the highest single-particle orbital that is occupied by the last nucleons. In reality, however, the situation is much complicated since the nuclear many-body system is strongly correlated due to the strong and singular short-range interaction between nucleons \cite{Talmi11,Dug12}. This is also true for states in magic nuclei which retain a significant single-particle character \cite{Dug12}.

\begin{figure}
\centerline{\includegraphics[width=0.45\textwidth]{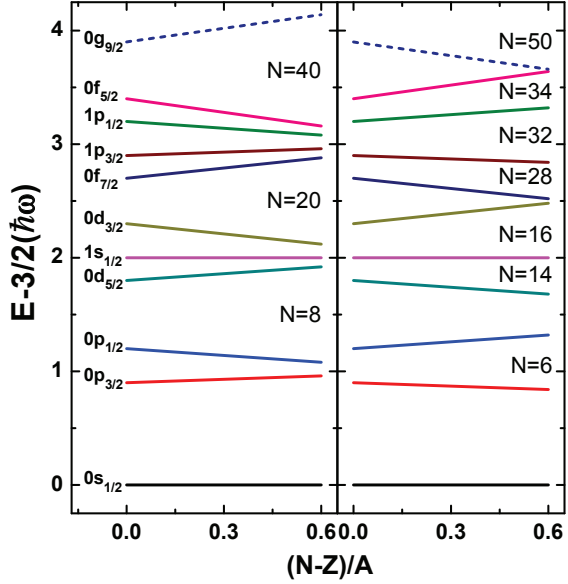}}
\caption{(Color online) The evolution of the shell structure as a function of $(N-Z)/A$ with the HO potential plus SO coupling of the form $\lambda(1+\kappa_{SO}\frac{N-Z}{A})\hbar\omega {\bf l}\cdot{\bf s}$ with $\lambda=0.2$ and $\kappa_{SO}=-1$ (left) and $1$ (right). The $0g_{9/2}$ orbital is shifted upwards by 0.3$\hbar\omega$ for a clearer presentation. \label{sch} }
\end{figure}

Atomic nuclei show striking regular features in spite of their complex nature. From a simple phenomenological point of view,  the shell structure is characterized by the presence of gaps in the calculated single-particle spectrum. As pointed out in 
Ref. \cite{Talmi11}, the shell model single-particle wave functions evaluated in this picture should be considered as model-dependent wave functions which may be very different from the real wave functions of the nucleus.
The HO mean field approximation with
SO coupling, which is known as the independent particle model, was the first successful model
to predict correctly the traditional magic numbers \cite{Goe49}. The calculated shell structure may change if an isospin dependence of the SO coupling is taken into account. To illustrate this point, in Fig. \ref{sch} we evaluate the HO single-particle spectra by adding an isospin-dependent SO coupling of the form 
$\lambda(1+\kappa_{SO}\frac{N-Z}{A})\hbar\omega {\bf l}\cdot{\bf s}$ \cite{Bohr69} to the HO potential. From this equation one sees that, if $\kappa_{SO}<0$, then the SO coupling gradually diminishes as $N-Z$ increases, i.e., approaching neutron-rich nuclei. In other words, as ones departs from $N=Z$ nuclei, the SO interaction has less and less importance and, therefore, traditional SO shell closures like $N=28$ and $N=50$ will disappear. In the limit, when the SO coupling vanishes completely, the spectra will be characterized by only HO magic numbers. This can be seen in the left panel of Fig. \ref{sch}. However, this contradicts most available experimental information on the shell evolution even though there is indication that the $N=28$ shell may have been eroded in $^{42}$Si \cite{Bas07,Win12,Uts12,Lap12,Force10}. On the other hand, a positive $\kappa_{SO}$ induces an enhancement of the SO coupling in neutron-rich nuclei. In such a case, the $N=8$ and 20 HO magic numbers will disappear while SO shell closures like $N=6$, $14$, 16, 32, 34 and 56 will appear, as seen in the right panel of Fig. \ref{sch}. It should be kept in mind, however, that the gaps at $N=28$ and 50 will be enhanced in neutron-rich nuclei within this naive picture.

In reality, we may expect that the SO coupling will be reduced in neutron-rich drip-line nuclei with a diffusive surface since the SO interaction is peaked at the nuclear surface. One also has to consider that the single-particle orbitals may show different $l$ dependence, depending upon the strength of the potential, which can lead to a systematic change of the shell structure \cite{Ham12}. Thus a more realistic description of the shell structure may be obtained with the Woods-Saxon potential
which has a glorious history of success and is still one of the most suitable models in describing the nuclear single-particle structure.
A variety of parameterizations of the Woods-Saxon potential exists (see, e.g., Refs. \cite{Bohr69,Blo60,Isa02,Dud82} and Table II in Ref. \cite{Sch07}). In the ``standard" one \cite{Bohr69,Blo60}, the strengths of the central and SO potentials are given as
\begin{equation}
V=V_0(1+\frac{4\kappa}{A}{\bf t}\cdot{\bf T}_{d}),
\end{equation}
and
\begin{equation}
V_{SO}=\lambda V_0(1+\frac{4\kappa_{SO}}{A}{\bf t}\cdot{\bf T}_{d}),
\end{equation}
where we have replaced the original $N-Z$ term with $4{\bf t}\cdot{\bf T}_{d}$ to get a consistent description of both protons and neutron orbitals.
${\bf t}$ and ${\bf T}_{d}$ denote the isospin quantum numbers of the last nucleon and of the daughter nucleus, respectively. The total isospin of the system is ${\bf T}={\bf t} + {\bf T}_{A-1}$.
It is $4{\bf t}\cdot{\bf T}_{A-1}=-3$ for the $T=0$ ground state of a $N=Z$ nucleus and 
\begin{eqnarray}
\nonumber 4{\bf t}\cdot{\bf T}_{A-1}&=&N-Z-1 {\rm ~for~neutron~orbits}\\
&=&-(N-Z+3)  {\rm ~for~proton~orbits~~~~}
\end{eqnarray}
in $N> Z$ nuclei with $T=(N-Z)/2$ \cite{Sch07}. In Ref. \cite{Bohr69}, the isospin-dependent terms in Eqs. (1) and (2) are parameterized as 
\begin{equation}
\kappa=\kappa_{SO}=-\frac{33}{51},
\end{equation}
where the SO potential
depth is assumed to have the same isospin dependence as that of the central
potential. This assumption is rather commonly used \cite{Dud82,Sch07}. The typical strength of $\kappa$ is in the range $-0.6\sim-0.9$.

\begin{figure}
\centerline{\includegraphics[width=0.48\textwidth]{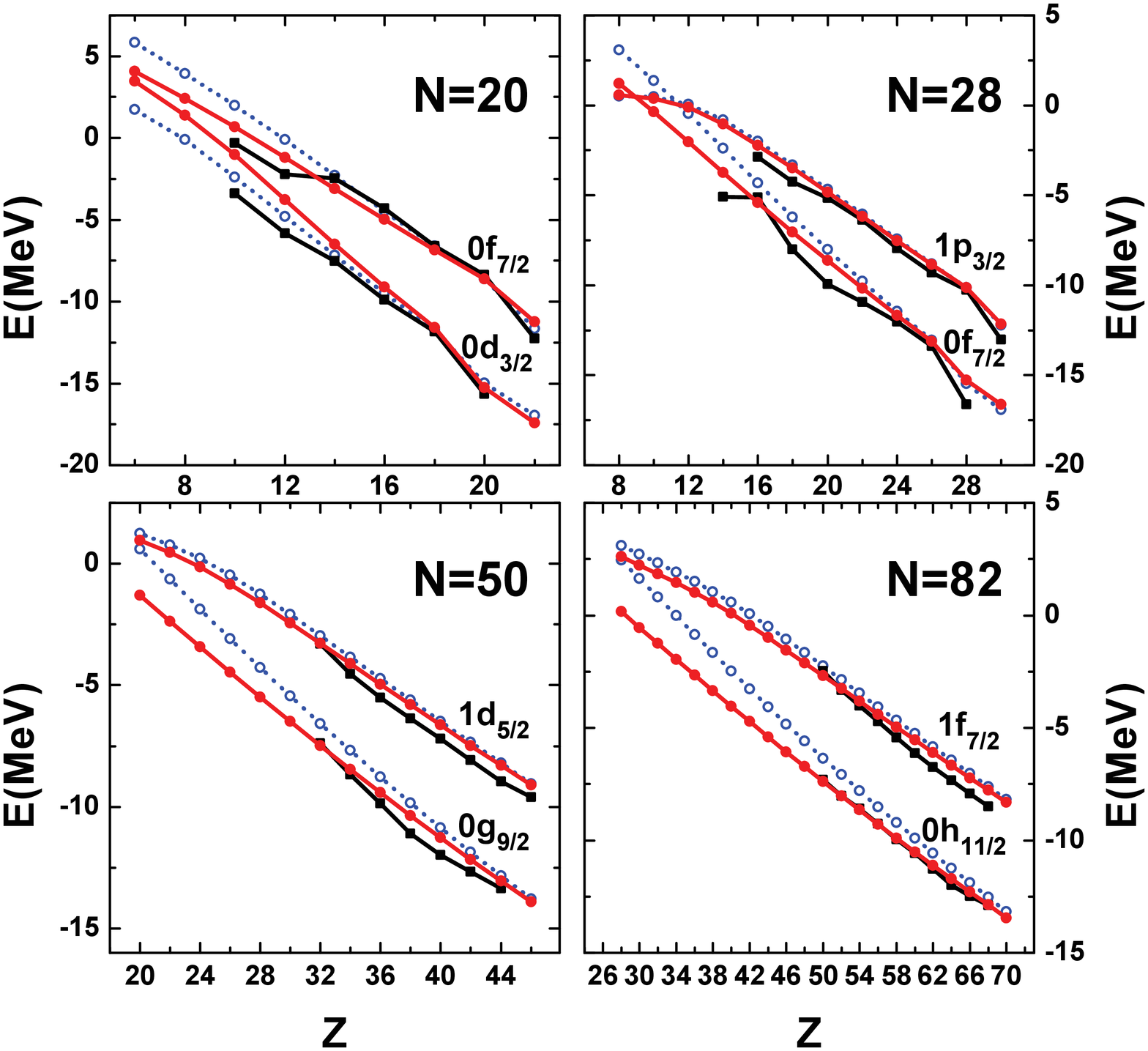} }
\caption{(Color online) Evolution of the N=20, 28, 50 and 82 gaps as a function of proton number $Z$ for calculations with the standard Woods-Saxon parameter and $\kappa_{SO}=\kappa$ (open circles) and $-\kappa$ (solid circles). The black squares denote the experimental one-neutron addition/removal energies \cite{nudat} of the nuclear states  with the same spin and parity as the corresponding calculated single-particle states  (see text for details).\label{stan}}
\end{figure}

We have done a systematic calculation on the shell evolution with the standard Woods-Saxon parameters (see, also, Ref. \cite{Ham12}). No correlation effect is considered at this stage. The calculated single-particle energies of neutron orbitals that are close to the $N=20$, 28, 50 and 82 shell closures are plotted in Fig. \ref{stan}.
As can be seen from the figure, such calculations suggest that SO shell closures like $N=28$, 50 and 82 will erode in neutron-rich nuclei but this seems to proceed too fast. Moreover, it cannot explain in a straight-forward way the disappearance of HO shell closures like $N=8$ and 20. The predicted $N=40$ gap is also too strong.
As we will show below, this problem may be fixed if we assume a strong positive $\kappa_{SO}$. We have found that calculations with other parameters \cite{Dud82,Sch07} will lead to the same conclusion. For comparison, in Fig. \ref{stan} we also plotted the experimental one-neutron addition/removal energies of the nuclear states with corresponding spin and parity. In general, however, it should be mentioned that the single-particle energy is not an observable. The energies calculated from mean-field models do not necessarily agree with measured nuclear energy levels, especially for those around middle shell that may be highly correlated \cite{Dug12}.

To get a qualitative idea on the role played by $\kappa_{SO}$, we performed two kinds of calculations using the standard Woods-Saxon parameter \cite{Bohr69} and taking $\kappa_{SO}=\kappa$ or $-\kappa$. The numerical code GAMOW was used \cite{Ver82}. Calculations on the evolution of the N=20, 28, 50 and 82 magic numbers are plotted in Fig. \ref{stan}.
The $N=20$ shell closure is expected to disappear in neutron-rich nuclei like $^{32}$Mg \cite{Sor08} (see, also, Ref. \cite{Wim10}). 
In calculations with the standard parameters, however, the gap in nuclei like $^{28}$O is as large as 4.2 MeV.
The $N=20$ shell persists even if one takes $\kappa_{SO}=0$, as shown in Ref. \cite{Sch07}. If one takes $\kappa_{SO}=-\kappa$, however, the $N=20$ gap in $^{28}$O would be reduced to only 1.2 MeV.

The situation around $N=28$ is somewhat complicated \cite{Bas07}. The $N=28$ gap will be reduced in neutron-rich nuclei as expected from the standard Woods-Saxon (WS) calculations \cite{Ham12}. But the reduction of this shell gap will be significantly retarded if a positive $\kappa_{SO}$ is assumed, as can be seen in Fig. \ref{stan}. 
Similarly, the $N=82$ shell closure may be reduced in very neutron-rich nuclei like $^{110}$Ni since the large $l$ orbital $0h_{11/2}$ will lose energy faster than other smaller-$l$ orbitals when the potential becomes shallower. Indeed, the distance between the $0h_{11/2}$ and $1f_{7/2}$ orbitals decrease from 4.1 MeV in $^{132}$Sn to 0.65 MeV in $^{110}$Ni in calculations with standard parameters. The gap increases to 2.5 MeV with $\kappa_{SO}=-\kappa$.
It should be mentioned that experimental data in Ni isotopes are available only up to $N=50$. The $1f_{7/2}$ orbital in $^{110}$Ni is predicted to be unbound in both calculations.

\begin{figure}
\centerline{\includegraphics[width=0.48\textwidth]{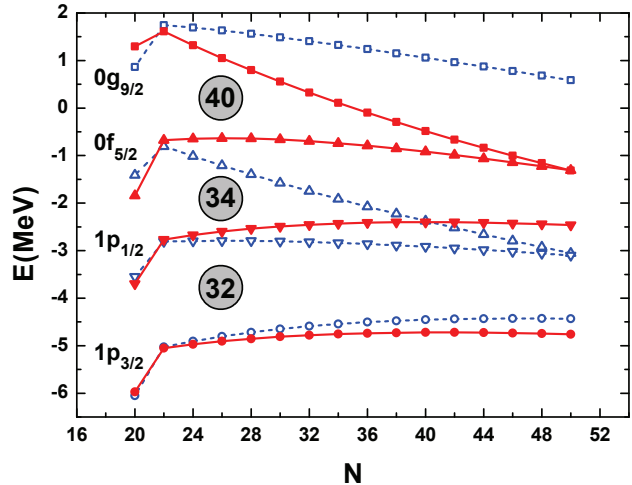} }
\caption{(Color online)  Evolution of the single-particle energies of the $1p_{3/2}$, $1p_{1/2}$, $0f_{5/2}$ and $0g_{9/2}$ orbitals in Ca isotopes as a function of the neutron number $N$ for calculations with the standard Woods-Saxon parameters and $\kappa_{SO}=\kappa$ (open symbols) or $\kappa_{SO}=-\kappa$ (solid symbols). In the former calculations, the $N=40$ shell gap is significantly enhanced in neutron-rich isotopes while the $N=34$ sub-shell gap is eroded. \label{ca}}
\end{figure}

The sign of $\kappa_{SO}$ can have significant influence on the relative strength of the $N=40$ and 50 shell gaps in neutron-rich nuclei. As examples, in Fig. \ref{ca} we plot the calculated single-particle spectra of Ca isotopes. The $N=40$ shell gap in $^{60}$Ca is predicted to be as large as 4.4 MeV in calculations with the standard parameters. While in the unbound nucleus $^{70}$Ca the $N=50$ gap is only 0.67 MeV.  The $N=50$ shell gap may disappear if the $0g_{9/2}$ orbital lose too much energy due to the shallowing of the central potential \cite{Ham12}. On the other hand, the $^{60}$Ca and $^{70}$Ca gaps are calculated to be
0.43 MeV and 2.3 MeV, respectively, if we take $k_{SO}=-k$, which means that a positive $k_{SO}$ will restore the $N=50$ magic number by ``destroying" the $N=40$ HO shell closure. However, there is no experimental indication that $N=40$ will emerge as a new magic number in neutron-rich nuclei \cite{Kan12,Nai12}. 

The $N=32$ gap between the orbitals $p_{3/2}$ and $p_{1/2}$ is not much affected by the $\kappa_{SO}$ term. The $N=32$ gap in $^{52}$Ca is calculated to be 1.8 MeV by using the standard WS parameters. But it increases to 2.3 MeV with $k_{SO}=-k$.
The energy difference between $1p_{1/2}$ and $0f_{5/2}$ is only 0.92 MeV in $^{54}$Ca. The $N=34$ gap increases to 1.7 MeV by taking $k_{SO}=-k$.

Recent experiments suggest that $^{22}$O and $^{24}$O should be doubly magic nuclei ~\cite{Sor08} (see, also, Refs. \cite{Hoff09,Kan09,Tsh12} for recent results). $^{24}$O is the heaviest bound oxygen isotope that has been observed so far. 
The $N=16$ gap between the neutron $1s_{1/2}$ and $0d_{3/2}$ states are measured to be $4.86\pm0.13$ MeV in Ref. \cite{Hoff09}. The values given by calculations with $\kappa_{SO}=\kappa$ and $-\kappa$ are 3.3 MeV and 4.5 MeV, respectively. The calculated $N=14$ gap in $^{20}$O increases from 1.8 MeV ($\kappa_{SO}=\kappa$) to $2.7$ MeV ($\kappa_{SO}=-\kappa$).

The $N=14$ gap disappears in C and N isotopes with nearly degenerate $0d_{5/2}$ and $1s_{1/2}$ orbitals. This is easily understood since the $0d_{5/2}$ orbital loses its energy faster when going towards the dripline, resulting in a nearly-degenerate $0d_{5/2}$ and $1s_{1/2}$ shells \cite{Ham12}. From a shell model point of view, where the above-mentioned mechanism is missing, this fact is related to the complicated interplay between the isovector and isoscalar two-body interactions \cite{Yuan12}.

A possible different form of  isospin dependence in the SO potential than that of the central potential has been the subject of several studies \cite{Isa02,Sch07,Kou00}. Ref. \cite{Sch07} assumed that $\kappa_{SO}=0$. The study of Ref. \cite{Isa02} showed that $\kappa_{SO}\sim 0.2$ to 0.7 can also explain the single-particle spectra in the neutron-rich nuclei $^{132}$Sn and $^{208}$Pb. They suggest that such an opposite value is consistent with the two-body SO interaction as well as the Walecka and Skyrme-Hartree-Fock calculations.
The study of Ref. \cite{Zal09} showed that the isoscalar SO coupling in the Skyrme force is reduced while the tensor coupling is strongly attractive, which may also indicate that the SO splitting can be enhanced in neutron-rich nuclei.
However, it should be mentioned that the spectra of heavy stable nuclei are quite insensitive to the sign of $\kappa_{SO}$ since the values of the term ${\bf t}\cdot{\bf T}_{d}/A $ are usually much smaller than those in light nuclei with large neutron excess. As a result, it is not determined by a normal global fitting procedure \cite{Kou00}.

\begin{table*}
  \centering
  \caption{Woods-Saxon potential parameters obtained by fitting to the available single-particle and single-hole states
around doubly-magic nuclei with the restriction $\kappa_{SO}=-\kappa$ and comparison with some existing parameters.}\label{table}
  \begin{tabular}{cccccccc}
  \hline 
\hline 
&$V_{0}$ (MeV)&$r_{0}$ (fm)&$r_{SO}$ (fm)&$a,a_{SO}$ (fm) &$\lambda$ &$\kappa$ &\\
\hline 
&50.92&1.285&1.146&0.691&24.07&0.644&$\kappa_{SO}=-\kappa$\\
Refs. \cite{Bohr69,Blo60}&51&1.27&1.27&0.67&32.13&0.647&$\kappa_{SO}=\kappa$\\
Ref. \cite{Dud82}&49.6&1.347(n)/1.275(p)&1.31(n)/1.32(p)&0.7&35(n)/36(p)&0.86&$\kappa_{SO}=\kappa$\\
Ref. \cite{Sch07}&52.06 &1.260 &1.16&0.662 &24.1 &0.639 &$\kappa_{SO}=0$\\
\hline 
\hline 
   \end{tabular}
\end{table*}

\begin{figure}
\centerline{\includegraphics[width=0.45\textwidth]{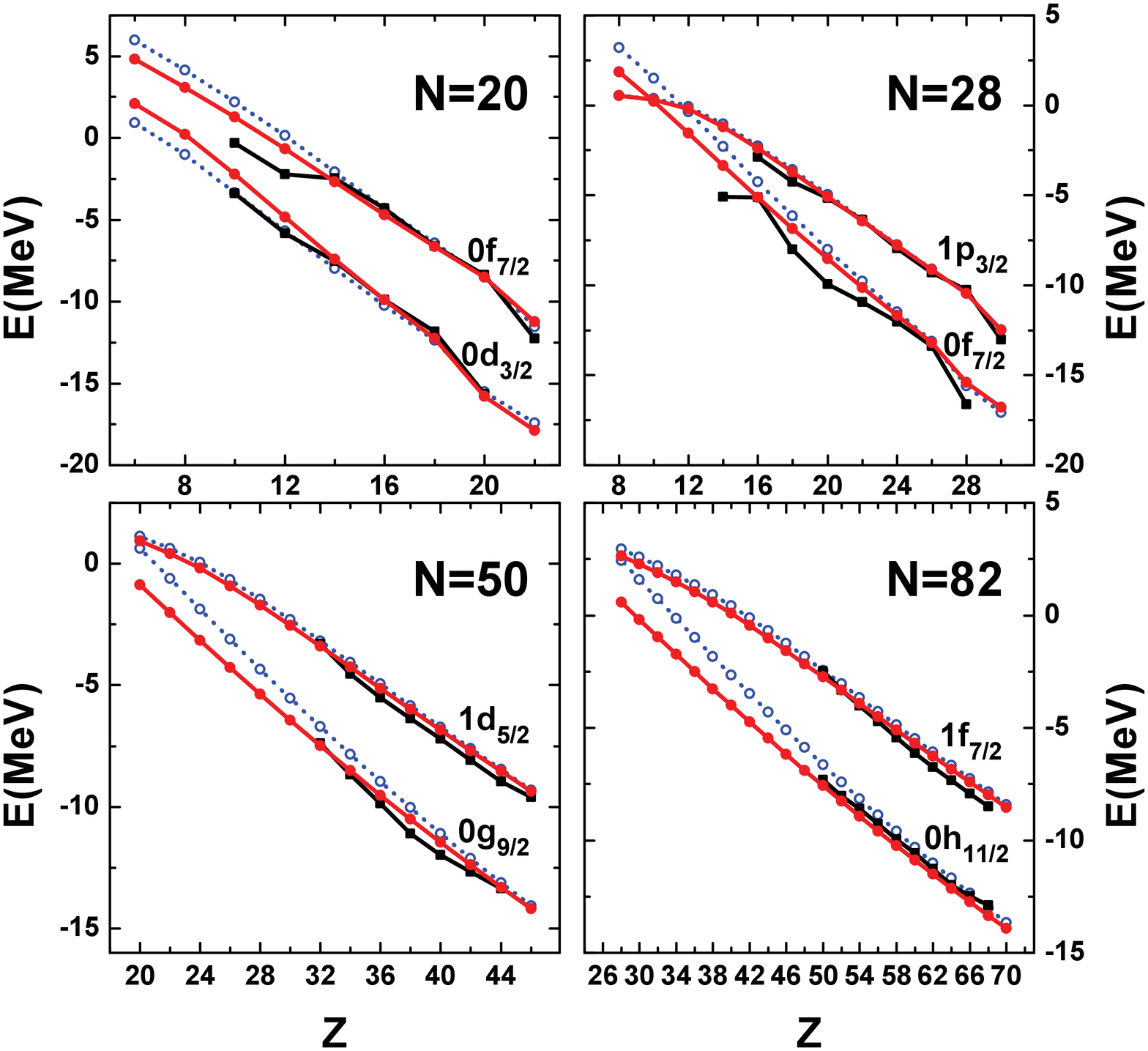} }
\caption{(Color online) Same as Fig. \ref{stan} but for calculations with the re-fitted parameter and $\kappa_{SO}=\kappa$ (open symbols) and $\kappa_{SO}=-\kappa$ (solid symbols). \label{fit}}
\end{figure}

The effect of the $\kappa_{SO}$ term may be partly swallowed by other Woods-Saxon parameters, in particular the SO radius parameter $r_{SO}$. $r_{SO}$ was taken as a free parameter in several calculations \cite{Dud82,Sch07}, with values which are smaller than that of the central potential, i.e., $r_0$. To explore this point further, we re-fitted the Woods-Saxon parameters under the restriction $\kappa_{SO}=-\kappa$.
The parameters of the Woods-Saxon potential are adjusted to single-particle and single-hole states
around the doubly-magic nuclei $^{16}$O, $^{40,48}$Ca,
$^{56}$Ni, $^{100}$Sn, $^{132}$Sn and $^{208}$Pb, as listed in Refs. \cite{Isa02,Sch07}. We use the same fitting procedure as employed in Ref. \cite{Qi12}. The fitting to nuclear single-particle energies usually favors larger values for the radius parameter $r_0$, which may lead to a bad description of nuclear charge radii and moments of inertia \cite{Wyss}. 
In this work we restrict that $r_0<1.3$ fm.
The results thus obtained are presented in Table \ref{table}.
Calculations on the evolution of the $N=20$, 28, 50 and 82 magic numbers are plotted in Fig. \ref{fit}, which show similar trends to those in Fig. \ref{stan}.
In this work we concentrated our attention on the structure of light nuclei. But calculations with the re-fitted parameter shown in Table \ref{table}  can describe the states in heavy magic nuclei equally well in comparison with those with the potentials given in Refs. \cite{Sch07,Dud82}. This is because in heavy nuclei the isospin dependent term in Eq. (2) are usually small and does not change much for a given isotopic or isotonic chain.

\begin{figure}
\centerline{\includegraphics[width=0.45\textwidth]{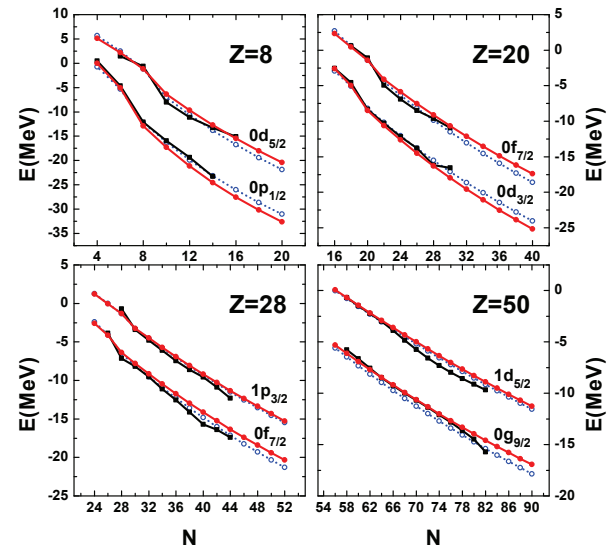}}
\caption{(Color online) Same as Fig. \ref{fit} but for calculations on the $Z=8$, 20, 28 and 50 proton shell gaps as a function of neutron number $N$ in neutron-rich nuclei. The open and solid symbols stand for calculations with $\kappa_{SO}=\kappa$ and $\kappa_{SO}=-\kappa$, respectively.\label{proton}}
\end{figure}

In Fig. \ref{proton} we show the influence of the isospin dependence in the SO coupling on the calculated proton single-particle spectra in neutron-rich nuclei. As seen from Eq. (3), a positive $\kappa_{SO}$ would suggest that the proton SO splitting is reduced in neutron-rich nuclei.
Ref. \cite{Sch04} does show that the splitting between the binding energies of the last proton in the lowest $7/2^+$ and $11/2^-$ states, which seem to have
consistent spectroscopic factors and exhibit near-single-particle-like
character, increases with neutron excess in neutron-rich Sb isotopes. As already shown 
in Ref. \cite{Bha06}, this fact can be reproduced if we take a positive value for $\kappa_{SO}$. Within the Skyrme-Hartree-Fock approach, the splitting is related the effect of the two-body tensor force  \cite{Colo07}.
It is difficult to draw any conclusion on the situation in light neutron-rich nuclei since experimental results are still inadequate. However, shell-model calculations tend to suggest that the splittings in the calculated shell model effective single-particle energies (centroid eigenvalues in relation to the monopole interaction) between SO partners like $0p_{3/2}$ and $0p_{1/2}$ and $0f_{7/2}$ and $0f_{5/2}$ are diminished with neutron excess \cite{Yuan12,Sie10}. This is consistent with the binding energy systematics in Ref. \cite{Sor08} which shows that the $Z=8$ and 20 gaps increase in neutron-rich nuclei. On the other hand, the $Z=28$ gap decreases in those nuclei \cite{Sor08,Fla09}. These results are consistent with our assumption that $\kappa_{SO}$ is positive. However, it should be mentioned that these quantities mentioned above may not be fully equivalent from a microscopic many-body point of view (see Ref. \cite{Dug12} for a detailed explanation).

\begin{figure}
\centerline{\includegraphics[width=0.48\textwidth]{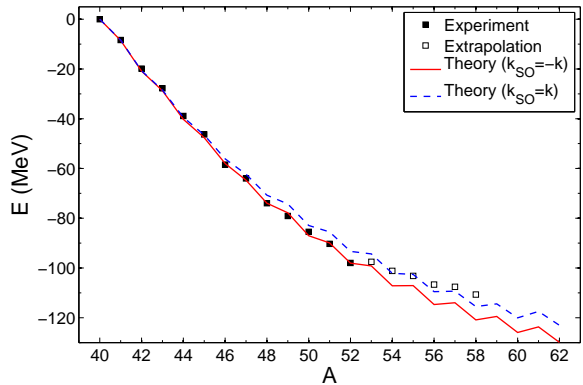} }
\caption{(Color online) Experimental \cite{Audi03,Gal12} and calculated ground-state energies of Ca isotopes, relative to that of $^{40}$Ca, as a function of mass number $A$.  \label{ca-2}}
\end{figure}

The single-particle scheme provides a zeroth-order approximation of nuclear structure which may be influenced by correlation effects including deformation, particle vibration coupling and pairing correlation \cite{Ben08,Del10,Tra96,Col10,Lit11} (see also Refs. \cite{Mah85,Ham74} for reviews on earlier calculations on the particle-vibration coupling in magic nuclei). As in Refs. \cite{Isa02,Sch07}, the particle-vibration coupling is not explicitly taken into account in our optimization of the parameters of the Woods-Saxon potential for simplicity. It is hoped that part of the effect of the particle-vibration coupling may be taken into account through the optimization of the parameters. A similar route is also used in some recent self-consistent mean field calculations (see, Refs. \cite{Ben08,Col10}, for further comments concerning this point).

To explore the influence of pairing correlation, we solved exactly the single-particle Woods-Saxon plus pairing Hamiltonian  
with a Lanczos diagonalization approach from Ref. \cite{Qi11}. We assumed that the ground states of even-even nuclei are all paired with seniority zero whereas those of odd-$A$ nuclei are assumed to be of seniority one.
We will present the semi-magic Ca isotopes as examples. These have been studied recently both experimental and theoretically \cite{Holt12,Gal12,Hag12,Cra10}. We performed calculations within a model space containing the $0f_{7/2}$, $0f_{5/2}$, $1p_{3/2}$, $1p_{1/2}$, $0g_{9/2}$ and $1d_{5/2}$ neutron orbitals by assuming $^{40}$Ca as the core.
To minimize the number of free parameters, a simple constant pairing strength, $G=1.795$ MeV, was employed in all calculations.
The ground state energies thus calculated are presented in Fig. \ref{ca-2} together with the corresponding experimental data. Calculations with $\kappa_{SO}=-\kappa$ predict larger binding energies for the neutron-rich nuclei $^{51-58}$Ca than the extrapolation values given in Ref. \cite{Audi03}. This is consistent with Ref. \cite{Gal12} where an additional binding is measured for the nuclei $^{51,52}$Ca.
The increased binding in Ca
isotopes may indicate a significant
subshell gap at $N=32$ \cite{Gal12}. There is a kink in the systematics of calculated two-neutron separation energies at $N=34$, which suggests that $N=34$ may also be a subshell. However, the $1p_{1/2}$ particle and $1p_{3/2}$ hole states are calculated to be nearly degenerate in the nucleus $^{53}$Ca even though there is a noticeable gap in the calculated single-particle spectrum (c.f., Fig. \ref{ca}).

\begin{figure}
\vspace{0.4cm}
\centerline{\includegraphics[width=0.47\textwidth]{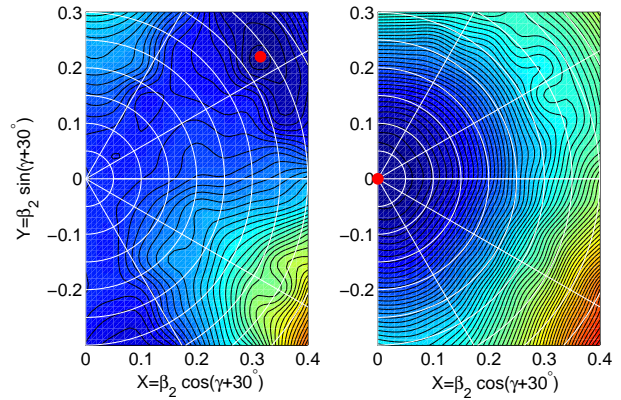}}
\caption{(Color online) Potential energy surface (PES) calculations for $^{32}$Mg ground state with the re-fitted Woods-Saxon parameter and $\kappa_{SO}=-\kappa$ (left) and $\kappa$ (right). For each $(\beta_2,\gamma)$ point the energy is minimized with respect to the $\beta_4$ deformation. The interval between neighboring contours is 0.1 MeV.\label{mg}}
\end{figure}

To analyze the influence of nuclear deformation on the shell evolution, we evaluated the potential energy surfaces of the nucleus $^{32}$Mg which has been intensively discussed recently \cite{Ots03,Yao11,For12,Hin11,Zhi06}. 
Relativistic and non-relativistic mean-field as well as Woods-Saxon  calculations  result in a spherical shape for the ground state of this nucleus \cite{Tera97,Rei99,Ben03}. This problem may be related to the fact that the predicted $N=20$ gap is rather large \cite{Yam04}. Whereas the observed large $B(E2)$ value for $^{32}$Mg indicates that the nucleus is strongly deformed with $\beta_2\sim0.5$ \cite{Mot95}. Our calculations using the re-fitted parameter of Table \ref{table} are plotted in Fig. \ref{mg}.
The minimum is around $\beta_2=0.38$ and $\gamma=4.8^{\circ}$ corresponding to the calculation with $\kappa_{SO}=-\kappa$.  It should also be mentioned that the shape of $^{32}$Mg is rather soft against $\beta_2$ deformation. The nucleus is calculated to be a rigid sphere with $\kappa_{SO}=\kappa$. Calculations with other parameters \cite{Bohr69,Dud82} lead to a similar conclusion.

\begin{figure}
\centerline{\includegraphics[width=0.48\textwidth]{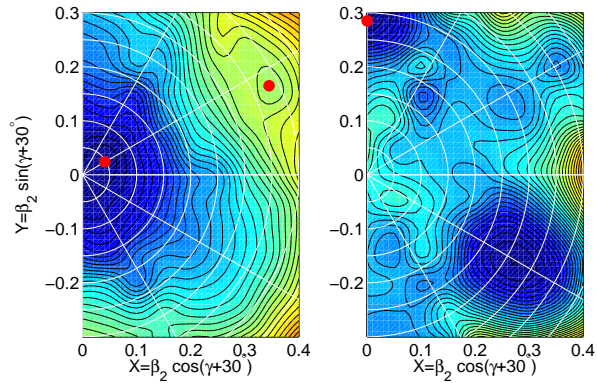} }
\caption{(Color online)  PES for the ground states in nuclei $^{34}$Si (left) and $^{42}$Si (right). \label{si}}
\end{figure}

For the calculations with $\kappa_{SO}=\kappa$ presented in Figs. 4-7, we have used the same Woods-Saxon parameters as those with $\kappa_{SO}=-\kappa$ in order to explore the effect of inversion of the sign of $\kappa_{SO}$. It should be mentioned that a quite similar result is obtained if we re-fit the Woods-Saxon parameters for calculations with  $\kappa_{SO}=\kappa$ or do the calculations with other parameter sets \cite{Bohr69,Dud82}. 

In a recent paper a second $0^+$ state in $^{34}$Si was observed, which shows a large deformation parameter of $\beta_2=0.29$ \cite{Rot12}. The ground state of this nucleus is calculated to be spherical with a coexisting shallow deformed minimum (or more exactly a shoulder). The calculated deformation of the second minimum is around $\beta_2=0.38$. 
This second minimum in $^{34}$Si disappears if one takes $\kappa_{SO}=\kappa$. This nucleus was also studied recently based on Hartree-Fock calculations \cite{Yao12,Nak12}.
The nucleus $^{42}$Si is calculated to be of oblate shape. This agrees with the shell model calculations with tensor force of Ref. \cite{Uts12} and the relativistic mean-field calculations in Ref. \cite{Li11}. Recent study on the ratio of the $4^+_1$ and $2^+_1$ energies in Si isotopes also indicate that the nucleus $^{42}$Si is characteristic of a well-deformed rotor \cite{Tak12}. The ground state in the $N=28$ isotone $^{44}$S is calculated to be of oblate shape with $\beta_2=0.27$ but the minimum is much shallower than that in $^{42}$Si. It is expected that in $^{44}$S both deformed and spherical
configurations coexist and mix weakly with
each other \cite{Force10}.

\begin{figure}
\centerline{\includegraphics[width=0.45\textwidth]{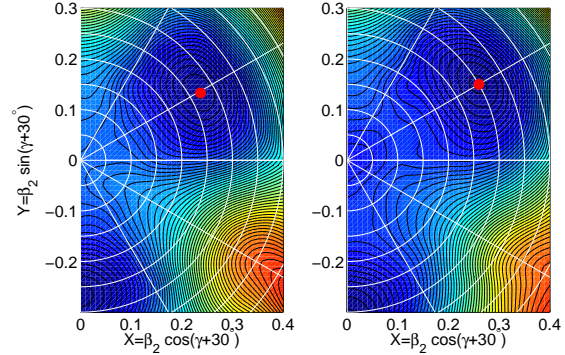} }
\caption{(Color online) PES for $N=40$ isotones $^{64}$Cr (left) and $^{66}$Fe (right). These nuclei are both calculated to be rigid spheres if we took  $\kappa_{SO}=\kappa$.\label{cr-fe}}
\end{figure}

It is expected that the quadrupole collectivity increases in neutron-rich nuclei around $N=40$ \cite{Len10,Bau12,Yang10,Sat12,hon09}. This is supported by our calculations with   $\kappa_{SO}=-\kappa$, as can be seen in Fig. \ref{cr-fe}. The calculated nuclear deformation in $^{64}$Cr is close to that given by the five-dimensional quadrupole collective Hamiltonian calculation in Ref. \cite{Sat12}. Such an enhanced collectivity also indicates that $N=40$ do not emerge as a magic number in neutron-rich nuclei.

In summary, we analyzed the shell structure of neutron-rich nuclei from a simple phenomenological single-particle point of view. We concentrated our attention on the attractive SO interaction since it determines the shell-model magic numbers.
We found that,
if the
SO splitting is relatively enhanced (i.e., with a strong positive value of $\kappa_{SO}$) in neutron-rich nuclei, both HO and Woods-Saxon calculations show that it will destroy the HO magic numbers $N=8$ and 20 and generate new SO magic numbers like $N=6$, $14$, 16, 32 and 34 instead. The traditional magic numbers $N=28$ and 50 will be eroded somehow in neutron-rich nuclei due to the sensitivity of larger-$l$ orbitals to the central potential depth. These SO shell closures are more robust than the HO magic numbers since their erosion is retarded by the relative enhancement of the SO splitting. In stable nuclei the $1d_{5/2}$ and $0g_{7/2}$ and the $0h_{9/2}$ and $1f_{7/2}$ orbitals are close to each other. These mechanisms both may split those shells, resulting in new shell closures like $N=56$ and 90. This is in agreement with the Skyrme-Hartree-Fock calculations with tensor force shown in Ref. \cite{Zal09}.
Such a simple scheme may give a quick estimation on the bulk properties of the single-particle spectra. It may also provide a convenient starting point for a variety of shell model calculations in the continuum (see, e.g., Ref. \cite{Xu11}) and  to explore the effect of the pairing correlation and deformation which may influence the shell structure.

We thank R. Liotta and R. Wyss for stimulating discussions and R. Wyss for bringing Ref. \cite{Isa02} to our attention when preparing the manuscript. CQ also thanks H.L. Liu for comments and S.A. Changizi for her help with the figures. 
This work has been supported by the Swedish Research Council (VR) under grant Nos. 621-2010-4723 and 621-2012-3805. ZX is supported in part by the China Scholarship Council under grant No. 2008601032.

\end{document}